\documentstyle[12pt,epsf]{article}

\newcommand{\beq}{\begin{equation}}
\newcommand{\eeq}{\end{equation}}
\newcommand{\beqa}{\begin{eqnarray}}
\newcommand{\eeqa}{\end{eqnarray}}
\newcommand{\noi}{\noindent}

\newcommand{\lsim}{\mathrel{\lower4pt\hbox{$\sim$}}
\hskip-12.5pt\raise1.6pt\hbox{$<$}\;}

\newcommand{\gsim}{\mathrel{\lower4pt\hbox{$\sim$}}
\hskip-12.5pt\raise1.6pt\hbox{$>$}\;}

\begin{document}
\begin{flushright}
AMES-HET-98-06\\
BNL-HET-98/19
\end{flushright}

\begin{center}
{\Large\bf Two body decays of the $b$-quark: \\
Applications to direct CP violation,\\ 
searches for electro-weak penguins and new\\ 
physics.}
\end{center}

\bigskip

\begin{center}
David Atwood$^{\rm a)}$ and Amarjit Soni$^{\rm b)}$
\medskip

{\small \it a) Department of Physics and Astronomy, Iowa State
University, Ames, IA\ \ 50011} 

{\small \it b) Theory Group, Brookhaven National Laboratory, Upton, NY\ \
11973} 
\end{center}

\begin{abstract}
A systematic experimental search for two-body hadronic decays of the
$b$-quark of the type $b\to$ 
quark${}+{}$meson is proposed. These reactions have a well defined
experimental signature and they should be theoretically cleaner
compared to exclusive decays. Many modes have appreciable
branching ratios and partial rate asymmetries may also
be quite large (about $8--50\%$) in several of them. 
In a few cases electroweak penguins appear
to be  dominant and may be measurable.  CP violating triple
correlation asymmetries provide a clean test of the Standard Model. 
\end{abstract}

%
%
Hadronic $B$-decays can be divided into two categories: inclusive and
exclusive. Inclusive decays involve calculations of quark level processes
and theoretical predictions here are relatively firm \cite{lenz}, but
well defined experimental signatures are usually rather difficult. On
the other hand, the theoretical calculations of exclusive channels
\cite{kramer,ali} entail many ad-hoc assumptions and crude
approximations \cite{atwoodone}. Consequently, the resulting
predictions are highly unreliable but the experimental signatures are
clear. In this paper we want to present a systematic study of a class
of semi-inclusive decays which lie somewhere in between the above two
categories. i.e.\ they possess improved prospects for predictive power
as well as well defined experimental signatures. Furthermore, we
suggest that the class of semi-inclusive processes that we will discuss,
1) should exhibit measurable direct CP  asymmetries in many channels, 2)
provide clean tests of the Standard Model (SM), 3) are a good probe of
electroweak penguins, and 4) should be helpful in clarifying
important issues related to hadronization.

We will focus here on two-body decays of the $b$-quark of the type:

\beq
b\to M+q_f \label{eqone}
\eeq

\noi where $M$ is a spin 0 or spin 1 meson and $q_f$ is a quark in the
final state. Note that two decays of the $b$-quark, that have received
considerable attention in the past few years, belong to this category;
namely $b\to \gamma s$ (i.e.\ $B\to \gamma X_x$) and $b\to \eta^\prime
s$ (i.e.\ $B\to \eta^\prime X_s$). Indeed the calculational procedure
that we will use for (\ref{eqone}) will be a generalization of the
method that we used, for $b\to \eta^\prime s$ leading to $B\to
\eta^\prime X_s$ \cite{atwoodtwo,atwoodthree,hou}. Two other cases for
which analogous calculations exist are $B\to \phi+X_s$
\cite{deshp} and $B\to K(K^\ast) +X$ \cite{browder}.

%
%

We first briefly discuss the experimental signature. There are a few
unique kinematic features of this class of events implied by~(\ref{eqone})
that should prove very helpful in searching for such processes:

\begin{enumerate}

\item The energy ($E_M$) of the meson $M$, will be centered around
$(m^2_b+m^2_M -m^2_{q_f})/2m_b$ and have a spread of $0(\Lambda_{\rm
QCD})\sim$ a few hundred MeV\null. 

\item The energy of the outgoing quark in (\ref{eqone}) is of course
also similarly fixed and since it is relatively low ($\sim2$ GeV), on
hadronization, it will lead to fairly low average multiplicity (about
3)/event. Thus the combinatorics problem in discriminating against
backgrounds is unlikely to be too difficult. This should be specially
helpful in discriminating against the cascade decays of charm hadrons as
charmless final states are of great interest for CP violation
\cite{lenz,kramer,bander,gerard}, for searching for electroweak
penguins (EWP) and for clues of new phyics 

\item In the rest frame of the $B$, the sum of the momentum transverse
to the direction of the meson $M$, over all the other particles in the
event, should be severely limited, perhaps to $0(\Lambda_{\rm QCD})\sim$
a few hundred MeV\null.

\end{enumerate}

We next want to outline the main reason why we believe that this class of
reactions will be
theoretically cleaner, compared to exclusive decays (say) into 2 mesons
\cite{kramer,ali}. The starting point for all such calculations is, of
course, the short distance Hamiltonian which can be symbolically
written as \cite{lenz}

\beq
H_{\rm eff} = \sum_j c_jO_j \label{eqtwo}
\eeq

\noi where $c$'s are the $c$-number coefficients and $O$'s the 4-quark
operators. Typically $O$'s have the form $\bar q_1\Gamma b\bar
q_2\Gamma^\prime q_3$ where $q$'s are the appropriate flavors of quarks,
$u$, $d$, $s$, $c$, and $\Gamma$'s are Dirac matrices and color indices
are suppressed for simplicity. The matrix elements leading from the
4-quark operators to reaction (\ref{eqone}) can be symbolically divided
into two categories

\beqa
\langle M_{q_2q_3} q_f |\bar q_1\Gamma b \bar q_2 \Gamma^\prime
q_3|b\rangle &\!\!\!\!\!\Rightarrow \langle M_{q_1q_2} |\bar q_2 \Gamma^\prime
q_3 |0\rangle \langle q_f|\bar q_1 \Gamma b|b\rangle &\qquad (a)
\nonumber \\
&\Rightarrow \langle M_{q_1q_3} | \bar q_1\Gamma^{\prime\prime} q_3
|0\rangle \langle q_f|\bar q_2\Gamma^{\prime\prime\prime} b|b\rangle
&\qquad (b) \label{eqthree} 
\eeqa

\noi where the subscripts on $M$ indicates the flavors of quarks that
$M$ is composed of.  For an explicit example consider the operator
$\bar u\gamma_\mu(1-\gamma_5) b \bar d \gamma^\mu (1-\gamma_5)u$
leading to the decays $b\to \rho^-u$ or $b\to \rho^0(\omega)d$
corresponding to eqns.~(\ref{eqthree}a) and (\ref{eqthree}b)
respectively.

Let us next consider the matrix element of the 4-quark operators in the
traditional exclusive reactions \cite{kramer,ali} $B\to M_1M_2$. It may
take the generic form:

\beq
\langle M_1M_2|\bar q_1 \Gamma b\bar q_2\Gamma^\prime q_3|B\rangle
\Rightarrow \langle M_1 |\bar q_2\Gamma^\prime q_3|0\rangle \langle
M_2|\bar q_1\Gamma b|B\rangle \label{eqfour}
\eeq

\noi Now recall that the evaluation of the factor $\langle M_2|\bar
q_1\Gamma b|B\rangle$ in eqn.~(\ref{eqfour}) requires a knowledge of two
form factors, (often denoted as $f_1$ and $f_0$) if $M_2$ is a $0^-$
meson or of four form factors ($A_{1\mbox{--}3}$, $V$) if $M_2$ is a
$1^-$ meson, at $(p_B-p_{M_2})^2 = m^2_{M_1}$. This represents the one
significant theoretical distinction between the semi-inclusive,
quasi-two-body reactions, eqn.~(\ref{eqone}), that we are considering in
this paper versus the exclusive decay $B\to M_1M_2$. 
%
%
The latter reaction, even using the factorization approach
of eqs.~(\ref{eqthree}-\ref{eqfour}), entails additional theoretical
uncertainties as it requires knowledge of 2--4 form factors 
for the $B\to M_2$ transition that the former reactions manage to
evade through summation over an appropriate ensemble of states.

In full generality the final states accessible through
eqn.~(\ref{eqone}) can be subdivided into three categories depending on
the type of operators which contribute: 1) tree${}\times{}$tree, 2)
tree${}\times{}$penguin and 3) penguin${}\times{}$penguin. There are
interesting physics issues that each type allows us to address.
However, for the purpose of this paper we will confine the discussion
to only those available via tree${}\times{}$penguin and
penguin${}\times{}$penguin. 

The results regarding the partial rate asymmetry (PRA) and the
branching ratio are presented in Table~\ref{tabone}\cite{tablenote}. 
A  figure of merit, often used to get a rough feel for detectability of CP
asymmetry, is given by $N^{3\sigma}_B$ defined by 

\beq
N^{3\sigma}_B = \frac{9}{\alpha^2_{\rm PRA} \cdot Br \cdot \epsilon_{\rm
eff}} \label{eqfive} 
\eeq

\noi where $N^{3\sigma}_B$ is the number of $B$-$\bar B$ pairs needed
to establish a PRA to the accuracy of three statistical standard
deviations. Here $\alpha_{\rm PRA}$ is the PRA, $Br$ is the branching
ratio and $\epsilon_{\rm eff}$ is the product of all the efficiencies
responsible for the signal. A quick calculation shows that with about
$5\times10^6 B$-$\bar B$ pairs, the asymmetries in the  $\rho^-$,
$K^-$ and $K^{\ast-}$ channels start to become accessible. With about
$5\times10^7 B$-$\bar B$, the PRA's in modes with $\pi^-$, $\pi^0$, $\rho^0$,
$\omega^0$, $D^-$ and $D^{\ast-}$ may also become measurable. Note also that
several channels may have 8--50\% asymmetries.

It is useful to recall that the existing CLEO analysis\cite{lingel}
is based on
about $3 \times 10^6$ $B$-$\bar B$ pairs and a factor of 2 to 2.5
times more data is expected to be analysed in the next few months. 
Thus even the existing data sample may well be sufficient to reveal
the PRA in some of the channels in the Table. Furthermore,
the $e^+ e^-$ based B-factories at Cornell, KEK and SLAC
are expected to have about a few times $10^7 
B$-$\bar B$ pairs/year starting next year. Thus PRA's in
many of these channels should be observable in the near future.

%
%

Using the CKM unitarity and assuming SU(3) it is easy to see that 
the difference in partial rates for $b\to \pi^-u (\rho^-u)$ are equal and
opposite to  that of $b\to K^-u(K^{\ast-}u)$. Thus, for instance,

\begin{eqnarray}
\Gamma(b\to \rho^-u) - \Gamma(\bar b\to \rho^+ \bar u) = -
\Gamma(b\to K^{*-}u) + \Gamma(\bar b\to K^{*+}\bar u)
\label{su3rel}
\end{eqnarray}

\noindent This implies that to the extent that $SU(3)$ is a valid
symmetry, in the standard model there will be no partial rate asymmetry in
$b\to h^- u$ where $h^-$ indicates the sum over $K$ ($K^*$) or $\pi$
($\rho$) final states. Conversely, if this combined asymmetry is very
large, it implies that physics beyond the standard model is present. From
the experimental point of view, it also illustrates that in order to see
asymmetries which might be present in the standard model, it is important
to be able to distinguish $\pi^-$ ($\rho^-$) from $K^-$ ($K^{-*}$).

The CP-violating PRA generated in these processes and given in
Table~\ref{tabone} originates from the absorptive part of the penguin graph
\cite{bander}. While traditionally most discussions of CP have centered
around the PRA thus obtained, we want to emphasize here that a very clean
test of the physics beyond the SM is possible by searching for triple
correlation asymmetry (TCA) via reaction (\ref{eqone}), specifically $b\to
qV$ where $V$ is a spin 1 meson. Notice first that in such decays a
CP-odd, TCA can be experimentally searched for by testing if
$\langle\sin\phi\rangle \ne0$, where $\phi$ is the azimuthal angle between
the decay plane of the vector $V$ and the plane of the leading two mesons
formed by $q$. 

We recall that TCA are $T_N$-odd observables which receive contributions
from real part of Feynman amplitudes i.e.\ they do not require the
presence of strong phases (unlike PRA). Thus they nicely complement tests
of CP violation that use $T_N$-even observables (e.g.\ PRA). However, for
a TCA \cite{notethat} to be present in $b\to qV$ there must exist a
corresponding TCA at 
the quark level. This can only happen if both left and right helicity
quark amplitudes are present with different CP-odd phases. The penguin
operators \cite{lenz} $O_5$ and $O_6$ do in fact contain coupling to
quarks of right helicity and since the tree operators (say $O_1$ and
$O_2$) couple to left-helicity quarks it would appear that the conditions
for TCA${}\ne0$ exist.  General considerations of helicity conservation, 
however, show this not to be the case for these operators if the final
meson is a vector. This is because a meson may only be constructed out of
a $q\bar q$ pair of the same chirality and since both of these operators
produce a $q\bar q$ pair of right handed chirality and a single $d$ or $s$
quark of left handed chirality, the right handed pair must bind to form
the meson and only the left handed $d$ or $s$ is left over in the final
state. The overall reaction thus has the same helicity structure as the
tree operator or the other penguins which only involve left chirality
quarks. Such amplitudes are thus suppressed by the ratios
of current quark masses: $m_s/m_b$ or $m_d/m_b$. For the case
when $M$ is a spin one meson \cite{spinless} TCA are, therefore, vanishingly
small  compared to PRA in the SM \cite{argu}. 

If such TCA's are detected beyond the level suggested by the 
the above suppressions, it indicates that some new physics  
generates effective operators with an odd number of right chirality 
quarks in the final state. Candidates would include models with right 
handed $W$ bosons, SUSY and some models with extended higgs sectors.




%
%

We have also examined the effect of EWP in this class of reactions. The
most interesting cases are those when the EWP contribution is not
color-suppressed where it is expected to be the largest (see Table).
Notice, in particular that for $b\to \pi^0 s$ and $\rho^0 s$ EWP
dominate over the other contributions. These should be observable with
$\sim5\times10^6 B$'s.
Again, the existing CLEO data sample may well be enough to reveal
the presence of EWP in these channels.

\begin{figure}
\caption{The 
normalized
recoil spectra for the
quasi-twobody decays, $b\to\pi^-u$ (solid), $K^{\ast-}u$ (dashed) and
$D^-c$ (dotted) are shown.
\label{figone}}
\epsfysize 7 in
\mbox{\epsfbox{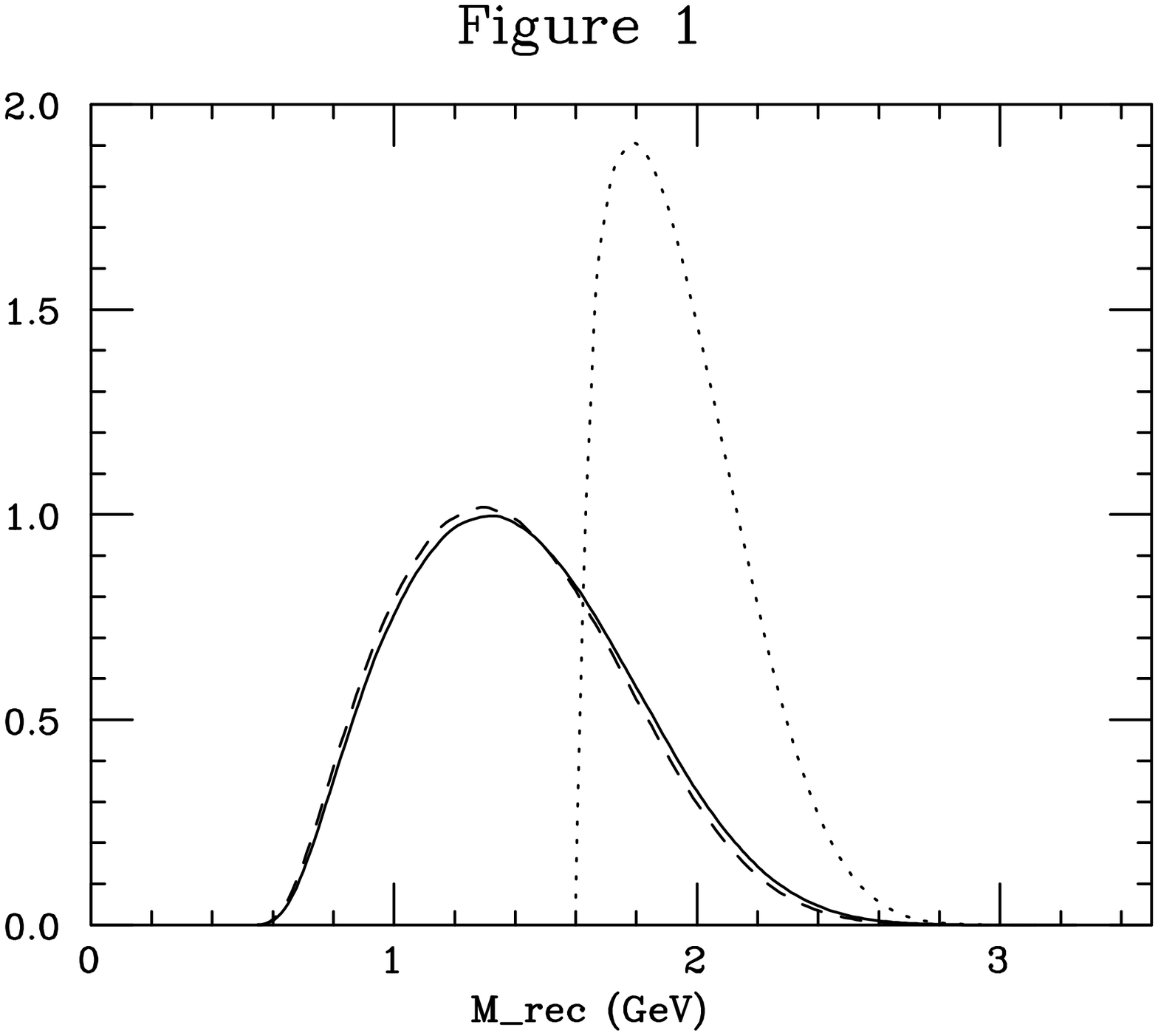}}
\end{figure}

It may be useful to understand the spectrum of the hadrons recoiling
against the meson $M$ in the $B$ decay. To do so we need to factor in
the Fermi motion of the $b$-quark with respect to the meson. For this
purpose we adopt the model of Ali and Greub \cite{alitwo} and use the
parameters chosen in the experimental analysis of $B\to \gamma X_s$
\cite{alam}. The resulting spectra are shown in Fig.~\ref{figone} for
a few cases. 

Finally it is important to note that in addition to the assumption of
factorization, both types of reactions, i.e.\ $b\to Mq_f$ and $b\to
M_1M_2$, require in eqns.~(\ref{eqthree}) and (\ref{eqfour})
respectively, a numerical value for $n_{eff}$, the effective number of
colors. In the calculations presented
in Table 1, we have tacitly {\it assumed\/} $n_{eff}=3$.
Hopefully, a more appropriate value for $n_{eff}$ can be
extracted from experiment after the branching ratios of a few of 
the modes in Table~\ref{tabone} are measured\cite{twentyone}. 
Thus the traditional
exclusive reactions ($B\to M_1M_2$) require the assumption of
factorization, a value for $n_{eff}$ and also the values of the form
factors whereas $b\to qV$ require only the first two. The key point is
that the problem of hadronic matrix elements calculations is so vastly
complicated that for the class of semi-inclusive 
reactions of eqn.~(\ref{eqone}), which  
require fewer assumptions and approximations, it may be
easier to extract information from experiment in an effort to
fine tune the calculational procedure for future applications.

To summarize, semi-inclusive decays of $B$ mesons emanating from
two-body decays of the $b$-quark have distinctive experimental
signature and they are theoretically cleaner compared to exclusive
decays. Moreover, as a rule,
this class of reactions have larger branching ratios~\cite{exception}
compared to exclusive channels and are expected to exhibit observable
CP-asymmetries with data samples $\gsim5\times10^6$ $B\bar B$
pairs. These reactions also provide good testing ground for the SM and
clues for new physics and are a good probe for electroweak penguins. It is
also important to note that since the underlying 
decays are that of a  $b$-quark, the corresponding reactions can be
studied using all types of $B$ mesons ($B_u$, $B_d$, $B_s$, $B_c$) and
at all kinds of $B$-facilities. Detailed studies of this class of $B$
decays could prove very helpful in clarifying many issues in QCD
dynamics governing weak decays which could in turn yield another route
to extraction of the CP violating CKM phases and tests of CKM unitarity. 
Systematic and dedicated
experimental searches for this class of modes are strongly encouraged.

\bigskip

\noi This research was supported in part by DOE contracts DE-AC02-98CH10886
(BNL) and DE-FG02-94ER40817 (ISU).

\bigskip

%
%

\begin{table}[ht]
\caption{Some modes of interest; $Br$, PRA and $N^{3\sigma}_B$ along with
EWP contributions (to color allowed channels only) are shown
\protect\cite{tablenote}. Note $\gamma\equiv\arg (-V^\ast_{ub}V_{ud}/V^\ast_{cb}
V_{cd})$.\label{tabone}} 
\begin{center}
\begin{tabular}{l|c|c|c|c}
\hline
Mode & $Br$ & $|{\rm PRA}|/\sin\gamma$(\%) 
& $N^{3\sigma}_B\sin^2\gamma \epsilon_{eff}/10^6$ & $Br$ due to EWP \\ 
\hline
$\pi^-u$           & $1.3\times10^{-4}$   & 8   & 12 &    \\
$\rho^-u$          & $3.5\times10^{-4}$   & 8   &  4 &    \\
$\pi^0d$           & $2.4\times10^{-6}$   & 36  & 28 &    
$4.7\times10^{-8}$ \\
$\rho^0d$          & $5.9\times10^{-6}$   & 38  & 10 &    
$1.3\times10^{-7}$ \\
$\omega d$        & $5.8\times10^{-6}$   & 39  & 10 &     
$7.0\times10^{-9}$ \\
$\phi d$           & $2.3\times10^{-7}$   & 0   &\  &     
$7.0\times10^{-9}$ \\
$K^0s$             & $2.5\times10^{-6}$   & 5   & 1200 &    \\
$K^{0^\ast}s$      & $2.9\times10^{-6}$   & 16  & 120 &    \\
$D^-c$             & $1.7\times10^{-3}$   & 2   & 17 &    \\
$D^{\ast-}c$       & $2.2\times10^{-3}$   & 2   & 13 &    \\
& & & \\
$K^-u$             & $2.9\times10^{-5}$   & 33  & 3 &    \\
$K^{\ast-}u$       & $5.1\times10^{-5}$   & 51  & 1 &    \\
$\bar K^0d$        & $2.0\times10^{-5}$   & 1   & 3000&    \\
$\bar K^{0^\ast}d$ & $2.6\times10^{-5}$   & 3   & 540 &    \\
$\pi^0s$           & $9.8\times10^{-8}$   & 0   &\  &$1.6\times10^{-6}$ \\
$\rho^0s$          & $2.5\times10^{-7}$   & 0   &\  &$4.3\times10^{-6}$ \\
$\omega s$       & $1.3\times10^{-6}$   & 0   &\  &$4.7\times10^{-7}$ \\
$\phi s$           & $6.3\times10^{-5}$   & 0   &\  &$4.7\times10^{-7}$ \\
$D^-_sc$           & $4.2\times10^{-2}$   & 0.1   & 300  &    \\
$D^{\ast-}_sc$     & $5.3\times10^{-2}$   & 0   & 300  &    \\ 
\hline
\end{tabular}
\end{center}
\end{table}
%
%
\clearpage


\begin{thebibliography}{99}
\bibitem{tablenote}
These results are obtained by using next-to-leading-order corrected
effective Hamiltonian; see \cite{buras,lenz, kramer}. We use
$m_c=1.4GeV$, $m_b=4.8GeV$,$|V_{ub}/V_{cb}|\approx 0.08$
and 
$|V_{td}/V_{ts}|\approx 0.22$
as in \cite{forty};
and also $m_s=150MeV$, $m_u=m_d=10MeV$.
The numbers shown are the central values obtained by using 
$\Lambda_{\overline{MS}}^{(5)}=225MeV$. Variations of $\Lambda$
over a 85MeV range changes the numerical values by about
$25\%$.  
More details will be given in a future publication.  
\bibitem{notethat} Note that triple correlation asymmetries are not
necessarily CP-odd. To establish the CP-odd nature may require
comparison of the asymmetry in $b$ and $\bar b$ decays.  
\bibitem{spinless} When $M$ is spinless, TCA do not vanish in the SM
even for massless $d$ or $s$ quarks; however, they are expected to be 
small. 
\bibitem{argu} We emphasize that this argument holds even in the
presence of electroweak penguins.  
\bibitem{twentyone}
It is quite possible that $n_{eff}$ is not the same for all the modes
in the Table. In particular, final states with $c \bar c$ in them
could have $n_{eff}$ different from other entries
in the Table.
\bibitem{exception}
Interesting exception to the rule is
the case when the tree contribution to $b \to q + M$
is color suppressed (e.g. $b \to \omega s$), 
and the corresponding 
exclusive two meson case is
color allowed (e.g. $B \to \omega K$). 
In such cases 
the presence of the spectator can make a large correction.
\end{thebibliography}
\end{document}